\documentclass[12pt]{article}
\usepackage{wallpaper}
\usepackage[margin=1in, paperwidth=8.5in, paperheight=11in]{geometry}
\usepackage{amsmath}
\usepackage{graphicx}%
\usepackage{amsfonts}%
\usepackage{amssymb}
\usepackage{color}
\usepackage{float,subfigure}
\usepackage[round]{natbib}
\usepackage{cite}
\usepackage{setspace}
\usepackage{comment}
\usepackage{enumerate}
\usepackage{verbatim}
\usepackage{rotating}
\usepackage{microtype}
\usepackage[normalem]{ulem}
\usepackage{url}
\usepackage[hidelinks]{hyperref} 
\usepackage{cancel}
\usepackage{animate}
\usepackage{mathtools}
\usepackage{epsfig}
\usepackage{epstopdf}

\usepackage{amsthm}

\hypersetup{pdfpagemode=UseNone}
\theoremstyle{plain}
\newtheorem{thm}{Theorem} 
\newtheorem{lem}{Lemma} 
\newtheorem*{thm*}{Theorem} 

\theoremstyle{definition}

\newcommand{\balpha}{ \mbox{\boldmath $ \alpha $} }
\newcommand{\bphi}{ \mbox{\boldmath $\phi$}}

\newcommand{\eps}{ \mbox{$\epsilon$}}

\newcommand{\btheta}{ \mbox{\boldmath $ \theta $} }
\newcommand{\sig}{ \ensuremath{\sigma}}

\newcommand{\blambda}{ \mbox{\boldmath $\lambda$} }

\newcommand{\ba}{ {\bf a} }

\newcommand{\bb}{ {\bf b} }

\newcommand{\bG}{ {\bf G} }

\newcommand{\bn}{ {\bf n} }

\newcommand{\bx}{ {\bf x} }

\newcommand{\by}{ {\bf y} }

\newcommand{\bz}{ {\bf z} }

\newcommand{\given}{\,\vert\,}

\newcommand{\MCAR}{\mbox{$\text{MCAR}$}}

\newcommand{\Pois}{\mbox{$\text{Pois}$}}

\newcommand{\NegBin}{\mbox{$\text{NegBin}$}}

\newcommand{\Gam}{\mbox{$\text{Gamma}$}}
\newcommand{\N}{\mbox{Norm}}
\newcommand{\Dir}{\mbox{Dir}}
\newcommand{\Mult}{\mbox{Mult}}

\begin{document}


\thispagestyle{empty}
\setcounter{page}{0}

\begin{center}
{\Large \textbf{Generating Poisson-Distributed Differentially Private Synthetic Data}} 

\bigskip

\textbf{Harrison Quick$^{*}$}\\ 
Department of Epidemiology and Biostatistics, Drexel University, Philadelphia, PA 19104\\
$^{*}$ \emph{email:} hsq23@drexel.edu

\end{center}

\textsc{Summary.}
The dissemination of synthetic data can be an effective means of making information from sensitive data publicly available while reducing the risk of disclosure associated with releasing the sensitive data directly. While mechanisms exist for synthesizing data that satisfy formal privacy guarantees, the utility of the synthetic data is often an afterthought. More recently, the use of methods from the disease mapping literature has been proposed to generate spatially-referenced synthetic data with high utility, albeit without formal privacy guarantees. The objective for this paper is to help bridge the gap between the disease mapping and the formal privacy literatures. In particular, we extend an existing approach for generating formally private synthetic data to the case of Poisson-distributed count data in a way that allows for the infusion of prior information.  To evaluate the utility of the synthetic data, we conducted a simulation study inspired by publicly available, county-level heart disease-related death counts.  The results of this study demonstrate that the proposed approach for generating differentially private synthetic data outperforms a popular technique when the counts correspond to events arising from subgroups with unequal population sizes or unequal event rates.


\newpage
\singlespacing
\doublespacing
\section{Introduction}
The Centers for Disease Control and Prevention's ``Wide-ranging Online Data for Epidemiologic Research'' (CDC WONDER) 
is a web-based tool 
for the dissemination of epidemiologic data collected by the National Vital Statistics System.  Via CDC WONDER, researchers can obtain detailed tables such as the number of deaths attributed to a specific cause of death (i.e., ICD code) in a given county and a given year by demographic variables such as age, race, and sex, subject to restrictions on small counts \citep{cdc:sharing}.  Unfortunately, such suppression techniques have been shown to be susceptible to certain types of targeted attacks \citep[e.g., ][]{dinur:nissim,holan2010,quick:zero}, thus motivating alternative methods for releasing public-use data with {formal} privacy {guarantees} with respect to the disclosure of sensitive information.

A popular approach for statistical disclosure limitation is the release of \emph{synthetic} data, as first proposed by \citet{rubin:1993} and \citet{little:1993}.  Specifically, if $\by=\left(y_1,\ldots,y_I\right)^T$ denotes a restricted-use dataset of $I$ potentially sensitive observations, a synthetic dataset, $\bz=\left(z_1,\ldots,z_I\right)^T$, can be generated by first fitting a statistical model, $p\left(\by\given \btheta\right)$, to the restricted-use data, obtaining the posterior distribution for the model's parameters, $\btheta$, and then generating $\bz$ from the posterior predictive distribution, $p\left(\bz\given \by\right) = \int p\left(\bz\given\by,\btheta\right) p\left(\btheta\given \by\right) d\btheta$.  When multiple synthetic datasets, $\bz^{(m)}$ for $m=1,\ldots, M$, are released, inference can then be made using the combining rules introduced by \citet{raghu:jerry} and \citet{reiter2003}, which allow the uncertainty due to imputation to be accounted for \citep{reiter:2002}.
Due to the flexible nature of producing synthetic data, models for data synthesis are often designed to accommodate complex data structures \citep[e.g.,][]{reiter2005,hu:BA,hu:jrssa}.  A recent example of this is the work of \citet{quick:synthetic}, which proposed the use of models from the disease mapping literature
to generate synthetic data, using ten years of stroke mortality data obtained from CDC WONDER as an illustrative example.  A more complete overview of synthetic data can be found in \citet{drechsler:text}.

While the risk of disclosure associated with the release of synthetic data is an active area of research \citep[e.g.,][]{reiter:mitra:2009,quick:smooth,hu:risk}, the drawback of many of the aforementioned approaches is the lack of \emph{formal} privacy guarantees, such as the concept of \emph{differential privacy} \citep{dwork:06}.  Specifically, if $\by$ denotes the true count data, a synthetic dataset $\bz$ is $\eps$-differentially private if for any hypothetical dataset $\bx=\left(x_1,\ldots,x_I\right)^T$ with $\Vert\bx-\by\Vert_1=2$ and $\sum_i x_i=\sum_i y_i$ --- {i.e., there exists $i$ and $i'$ such that $x_i=y_i-1$ and $x_{i'}=y_{i'}+1$ with all other values equal} --- then \begin{align}
\left\vert \log \frac{p\left(\bz\given \by,\btheta\right)}{p\left(\bz\given \bx,\btheta\right)}\right\vert \le \epsilon.\label{eq:dp_dfn}
\end{align}
While $\btheta$ can be viewed as a vector of model parameters, in practice the elements of $\btheta$ are specified in order to satisfy $\eps$-differential privacy.  While it would be impossible to exhaustively list the various mechanisms designed to satisfy~\eqref{eq:dp_dfn} --- though \citet{bowen:liu} provides an excellent review --- many are based on adding noise from a Laplace \citep{dwork:etal}, exponential \citep{mcsherry:talwar}, or geometric distribution \citep{ghosh:geometric}. Properties of differentially private mechanisms from a statistical prospective are discussed by \citet{wasserman:zhou}.

The first production system to use differential privacy was the Census Bureau's OnTheMap --- a mapping program for disseminating information about commuting patterns in the United States --- which was based on the framework proposed by \citet{onthemap}.  In particular, the data underlying OnTheMap are based on individual-level pairs of origin and destination Census blocks; for each destination block, \citet{onthemap} modeled the number of people commuting from each of the roughly 8 million Census blocks using a multinomial likelihood with a Dirichlet prior.  The authors then demonstrated that when this prior was sufficiently informative, synthetic data generated from the posterior predictive distribution would satisfy $\eps$-differential privacy.
In addition to OnTheMap, differentially private methods have been implemented by Google \citep{dp:google}, Apple \citep{dp:apple}, and Microsoft \citep{dp:microsoft}.
Moreover, the U.S.\ Census Bureau recently announced \citep{census:dp} that the 2018 End-to-End Census Test would be protected using differential privacy with an eye toward its use for the full 2020 Census.  A discussion of the challenges this process has entailed is provided by \citet{garfinkel:2018}.

While data on CDC WONDER \emph{can} be thought of in terms of a contingency table (e.g., $\text{county} \times \text{age-group}$) with a multinomial distribution where the goal would be to estimate the probability of an event occurring in a given cell, it's more common in the disease mapping literature to model the counts using a Poisson distribution \citep[e.g.,][]{brillinger} where the goal would be to make inference on the group-specific mortality rates at the county level.  For instance, while \citet{clayton:kaldor} represents an early, empirical Bayesian approach, the conditional autoregressive (CAR) model of \citet{bym} and its multivariate extension --- the multivariate CAR (MCAR) of \citet{gelfand:mcar} --- have served as the basis for fully Bayesian advances in spatial statistics in recent years
\citep[e.g.,][]{jon,datta:dagar,nonsep:poisson}.

The objective for this paper is to help bridge the gap between the disease mapping and differential privacy literatures.  Whereas \citet{quick:synthetic} proposed the use of standard disease mapping models to generate synthetic data without formal privacy protections, our goal here is to extend the formal privacy protections introduced by \citet{onthemap} to the setting of Poisson-distributed count data.
Full details of the methods explored in this paper are described in Section~\ref{sec:methods} --- this includes both background information regarding the multinomial-Dirichlet model proposed by \citet{onthemap} and the Poisson-gamma model proposed here.
To compare and contrast these two approaches, we have conducted a simulation study in Section~\ref{sec:sim} based on heart disease mortality data from U.S.\ counties.  In particular, we will explore the effect of heterogeneity in population sizes and the underlying event rates on the utility of the synthetic data produced by these approaches.
We then provide concluding remarks and discuss avenues for future research in this area.

\section{Methods}\label{sec:methods}
For the following presentation, we let $y_i$ denote the number of events belonging to group $i$ out of a population of size $n_i$, for $i=1,\ldots, I$ and $I\ge 2$.  While individual $y_i$ is deemed potentially sensitive, we assume $y_{\cdot}=\sum_i y_i>0$ is not sensitive and thus is publicly available.
\subsection{Multinomial-Dirichlet model}\label{sec:md}
To model $\by$, one option is to assume
$\by\given \btheta \sim \Mult\left(y_{\cdot}, \btheta\right)$ and further assume that $\btheta \sim \Dir\left(\balpha\right)$, where $\balpha=\left(\alpha_1,\ldots,\alpha_I\right)^T$ is a vector of hyperparameters to be defined shortly.  To generate a synthetic data vector, $\bz=\left(z_1,\ldots,z_I\right)^T$, with a given $\sum_i z_i=z_{\cdot}=y_{\cdot}$, one can first draw a sample, $\btheta^*$, from the posterior distribution for $\btheta$ --- i.e., $\btheta\given \by \sim \Dir\left(\by+\balpha\right)$ --- and then sample $\bz$ from the posterior predictive distribution, $p\left(\bz\given \by,\btheta,\balpha\right)$, by sampling from $\bz\sim \Mult\left(z_{\cdot},\btheta^*\right)$.  This is equivalent to integrating $\btheta$ out of the model and sampling $\bz$ from
\begin{align}
p\left(\bz\given\by,\balpha\right) =& \int p\left(\bz\given \btheta,\balpha\right) \times p\left(\btheta\given\by,\balpha\right) d\btheta\notag\\
=& \int \frac{z_{\cdot}!}{\prod z_i !} \times \prod \theta_i^{z_i} \times \frac{\Gamma\left(\sum y_i+\alpha_i\right)}{\prod \Gamma\left(y_i+\alpha_i\right)} \times \prod \theta_i^{y_i+\alpha_i-1} d\btheta\notag\\
=& \frac{z_{\cdot}!}{\prod z_i !} \times \frac{\Gamma\left(\sum y_i+\alpha_i\right)}{\prod \Gamma\left(y_i+\alpha_i\right)} \times \int \prod \theta_i^{z_i+y_i+\alpha_i-1} d\btheta\notag\\
=& \frac{z_{\cdot}!}{\prod z_i !} \times \frac{\Gamma\left(\sum y_i+\alpha_i\right)}{\prod \Gamma\left(y_i+\alpha_i\right)} \times \frac{\prod \Gamma\left(z_i+y_i+\alpha_i\right)}{\Gamma\left(\sum z_i+y_i + \alpha_i\right)}.\label{eq:multpost}
\end{align}
For the data synthesizer in~\eqref{eq:multpost} to satisfy $\eps$-differential privacy, we must satisfy the definition in~\eqref{eq:dp_dfn}
--- i.e., we require
\begin{align}
\left\vert\log\left(\frac{p\left(\bz\given\by,\balpha\right)}{p\left(\bz\given\bx,\balpha\right)}\right)\right\vert = \left\vert \log\left(\frac{\prod \Gamma\left(\alpha_i+x_i\right)}{\prod \Gamma\left(\alpha_i+y_i\right)} \times \frac{\prod \Gamma\left(z_i+\alpha_i+y_i\right)}{\prod \Gamma\left(z_i+\alpha_i+x_i\right)}\right)\right\vert \le \epsilon, \label{eq:multratio1}
\end{align}
for an $I$-vector of hypothetical data, $\bx$, such that $\Vert\bx-\by\Vert_1=2$ and $\sum_i x_i = \sum_i y_i$.
Without loss of generality, we assume the only differences in $\bx$ and $\by$ exist between the pairs $\left(x_1,x_2\right)$ and $\left(y_1,y_2\right)$ and furthermore that $x_1=y_1-1$ and $x_2=y_2+1$.  This implies that the expression in~\eqref{eq:multratio1} can be further simplified as
\begin{align}
\frac{p\left(\bz\given\by,\balpha\right)}{p\left(\bz\given\bx,\balpha\right)}
&=
\frac{\alpha_1+y_1}{\alpha_2+y_2-1} \times
\frac{z_2+\alpha_2+y_2-1}{z_1+\alpha_1+y_1}.\label{eq:multratio2}
\end{align}
We now wish to maximize and minimize~\eqref{eq:multratio2} for $z_1+z_2\le z_{\cdot}$.  To maximize~\eqref{eq:multratio2}, we let $z_1=0$ and $z_2=z_{\cdot}$, which implies that~\eqref{eq:multratio2} is maximized when $y_2=1$.  Similarly, to minimize~\eqref{eq:multratio2}, we let $z_2=0$ and $z_1=z_{\cdot}$, which implies that~\eqref{eq:multratio2} is minimized when $y_1=0$.  That is,
\begin{align*}
\frac{\alpha_1}{z_{\cdot}+\alpha_1} \le \frac{p\left(\bz\given\by,\balpha\right)}{p\left(\bz\given\bx,\balpha\right)} \le \frac{z_{\cdot}+\alpha_2}{\alpha_2},
\end{align*}
and thus to satisfy $\epsilon$-differential privacy, we require
\begin{align}
\epsilon = \log \frac{z_{\cdot}+\min \alpha_i}{\min \alpha_i} &\implies \min \alpha_i \ge \frac{z_{\cdot}}{\exp\left(\eps\right)-1}. \label{eq:dp_req_md}
\end{align}

As discussed in \citet{onthemap}, the restriction on $\balpha$ in~\eqref{eq:dp_req_md} is often overly strict.  For instance, suppose we wish to synthesize $z_{\cdot}=\text{10,000}$ events and allocate them across the approximately $I=\text{3,000}$ counties in the U.S.  If we let $\eps=7$, the result from~\eqref{eq:dp_req_md} would require \emph{each} $\alpha_i\ge 9.12$.  Considering that this is nearly three times the average number of events per county ($\text{10,000}\slash\text{3,000}=3.33$) ---
and considering that \citet{dwork:06} recommends selecting $\eps<1$ --- it would seem that~\eqref{eq:dp_req_md} requires us to use \emph{very} informative priors to achieve even modest levels of differential privacy. Furthermore, as the probability of generating a synthetic dataset with \emph{all} of the events in a single cell --- i.e., $z_i=z_{\cdot}$ and $z_{i'}=0$ for $i'\ne i$ --- is \emph{extremely} low, our concern for such extreme scenarios may be misplaced.  As a result of this limitation of pure differential privacy, \citet{onthemap} proposed a relaxed definition of differential privacy referred to as \emph{$\left(\eps,\delta\right)$-probabilistic differential privacy} in which 
a synthesizer 
satisfies $\eps$-differential privacy with probability $1-\delta$ for $\eps,\delta>0$.
While an $\left(\eps,\delta\right)$-probabilistic differentially private synthesizer will produce data with greater utility, an alternative that satisfies pure $\eps$-differential privacy (i.e., $\delta=0$) will be discussed in Section~\ref{sec:disc}. 

\subsection{Poisson-gamma synthesizer}\label{sec:pg}
A key drawback of generating synthetic data from the model in~\eqref{eq:multpost} is that, \emph{a priori}, each individual event has an equal probability of being assigned to any group.  This ignores potential heterogeneity in group-specific population sizes and geographic variation in event rates.  To address these limitations, we instead consider the case where
\begin{align}
y_i\given\lambda_i\sim \Pois\left(n_i\lambda_i\right) \;\;\text{and}\;\;\lambda_i\sim \Gam\left(a_i,b_i\right),\label{eq:pglik}
\end{align}
where $\lambda_i$ denotes the event rate in group $i$ and $a_i$ and $b_i$ denote group-specific hyperparameters.  In particular, we can consider $a_i$ a measure of the informativeness of the gamma prior in~\eqref{eq:pglik} and use $b_i$ to control $E\left[\lambda_i\right]=a_i\slash b_i$; the default choice would be to let $E\left[\lambda_i\right]=y_{\cdot}\slash n_{\cdot}$ --- the overall average rate --- and thus let $b_i=a_i n_{\cdot}\slash y_{\cdot}$.  We can also infuse prior information into our specification of $E\left[\lambda_i\right]$.  As we will illustrate in Section~\ref{sec:sim}, this can improve the utility of our synthetic data, though as discussed in Section~\ref{sec:disc}, this has implications with respect to the privacy budget.
Using Bayes Theorem, it is straightforward to show that
\begin{align}
\lambda_i\given y_i \sim \Gam\left(y_i+a_i,n_i+b_i\right),\label{eq:lambdapost}
\end{align}
and thus that
\begin{align}
p\left(z_i\given y_i,a_i,b_i\right) &= \int \Pois\left(z_i\given n_i\lambda_i\right) \times 
\Gam\left(\lambda_i\given y_i+a_i,n_i+b_i\right) d\lambda_i\notag\\
&=\frac{\Gamma\left(z_i+y_i+a_i\right)}{z_i! \times \Gamma\left(y_i+a_i\right)} \left(\frac{n_i}{b_i+2*n_i}\right)^{z_i} \left(\frac{b_i+n_i}{b_i+2*n_i}\right)^{y_i+a_i},\label{eq:znegbin}
\end{align}
which implies $z_i\given y_i,a_i,b_i \sim \NegBin\left(y_i+a_i,n_i\slash \left(b_i+2*n_i\right)\right)$.
While the posterior predictive distribution for $z_i$ in~\eqref{eq:znegbin} is sufficient for synthesizing a collection of independent, unconstrained $z_i$, we desire the distribution for $\bz$ \emph{conditioned} on $\sum_i z_i=z_{\cdot}$.  Without loss of generality, if we restrict our focus to the case where $I=2$ (i.e., region $i$ versus \emph{not} region $i$),
the distribution we desire is instead
\begin{align}
p\left(\bz\given \by,\ba,\bb,z_{\cdot}\right)
&=\frac{
\frac{\Gamma\left(z_1+y_1+a_1\right)}{z_1!} \left(\frac{n_1}{b_1+2*n_1}\right)^{z_1} \times
\frac{\Gamma\left(z_2+y_2+a_2\right)}{z_2!} \left(\frac{n_2}{b_2+2*n_2}\right)^{z_2}
}{\sum_{z=0}^{z_{\cdot}} \frac{\Gamma\left(z+y_1+a_1\right)}{z!} \left(\frac{n_1}{b_1+2*n_1}\right)^{z} \times
\frac{\Gamma\left(z_{\cdot}-z+y_2+a_2\right)}{\left(z_{\cdot}-z\right)!} \left(\frac{n_2}{b_2+2*n_2}\right)^{\left(z_{\cdot}-z\right)} }. \label{eq:poiscond}
\end{align}
Unfortunately, 
further simplification of the denominator in~\eqref{eq:poiscond} appears non-trivial, as demonstrated by Lemma~\ref{thm:den}:
\begin{lem}\label{thm:den}
Let $z_{\cdot}$, $c_1$, and $c_2$ be {positive integers} and let $p>0$ and $q>0$.  Then
\begin{align*}
\sum_{z=0}^{z_{\cdot}} \frac{\Gamma\left(z+c_1\right)}{z!} \times
\frac{\Gamma\left(z_{\cdot}-z+c_2\right)}{\left(z_{\cdot}-z\right)!} p^{z} q^{\left(z_{\cdot}-z\right)} =  \frac{d^{c_1-1}}{dp^{c_1-1}} \frac{d^{c_2-1}}{dq^{c_2-1}}  \frac{p^{c_1-1} q^{z_{\cdot}+c_2}-p^{z_{\cdot}+c_1}q^{c_2-1}}{q-p}.\end{align*}
\end{lem}
\noindent
Thus, when $c_1=y_1+a_1$ and/or $c_2=y_2+a_2$ are large, closed-form expressions for the denominator in~\eqref{eq:poiscond} may not be tractable, complicating our ability to specify criteria for $\ba$ and $\bb$ that will result in an $\eps$-differentially private $\bz$.  
See Appendix~A for a proof of Lemma~\ref{thm:den}.

\subsubsection{Requirements to satisfy differential privacy}
To satisfy $\eps$-differential privacy, we need to evaluate the ratio $p\left(\bz\given\by,\ba,\bb,z_{\cdot}\right)\slash p\left(\bz\given \bx,\ba,\bb,z_{\cdot}\right)$ where $\bx=\left(x_1,x_2\right)^T$ represents a set of hypothetical data such that $\left\Vert\bx-\by\right\Vert_1=2$; i.e.,
\begin{align}
\frac{p\left(\bz\given\by,\ba,\bb\right)}{p\left(\bz\given \bx,\ba,\bb\right)} =& \frac{\frac{
\frac{\Gamma\left(z_1+y_1+a_1\right)}{z_1!} \left(\frac{n_1}{b_1+2*n_1}\right)^{z_1} \times
\frac{\Gamma\left(z_2+y_2+a_2\right)}{z_2!} \left(\frac{n_2}{b_2+2*n_2}\right)^{z_2}}{\sum_{z=0}^{z_{\cdot}} \frac{\Gamma\left(z+y_1+a_1\right)}{z!} \left(\frac{n_1}{b_1+2*n_1}\right)^{z} \times
\frac{\Gamma\left(z_{\cdot}-z+y_2+a_2\right)}{\left(z_{\cdot}-z\right)!} \left(\frac{n_2}{b_2+2*n_2}\right)^{z_{\cdot}-z}}
}{\frac{
\frac{\Gamma\left(z_1+x_1+a_1\right)}{z_1!} \left(\frac{n_1}{b_1+2*n_1}\right)^{z_1} \times
\frac{\Gamma\left(z_2+x_2+a_2\right)}{z_2!} \left(\frac{n_2}{b_2+2*n_2}\right)^{z_2}}{\sum_{z=0}^{z_{\cdot}} \frac{\Gamma\left(z+x_1+a_1\right)}{z!} \left(\frac{n_1}{b_1+2*n_1}\right)^{z} \times
\frac{\Gamma\left(z_{\cdot}-z+x_2+a_2\right)}{\left(z_{\cdot}-z\right)!} \left(\frac{n_2}{b_2+2*n_2}\right)^{z_{\cdot}-z}}}\notag\\
=& \frac{C\left(\bx,\bn,\ba,\bb,z_{\cdot}\right)}{C\left(\by,\bn,\ba,\bb,z_{\cdot}\right)}
\times \frac{\Gamma\left(z_1+y_1+a_1\right)}{\Gamma\left(z_1+x_1+a_1\right)}
\frac{\Gamma\left(z_2+y_2+a_2\right)}{\Gamma\left(z_2+x_2+a_2\right)},\label{eq:pgdp}
\end{align}
where
\begin{align}
C\left(\by,\bn,\ba,\bb,z_{\cdot}\right) = \sum_{z=0}^{z_{\cdot}} \frac{\Gamma\left(z+y_1+a_1\right)}{z!}\frac{\Gamma\left(z_{\cdot}-z+y_2+a_2\right)}{\left(z_{\cdot}-z\right)!} \times r_1\left(\bn,\bb\right)^{z},\label{eq:pg_norm}
\end{align}
and $r_i\left(\bn,\bb\right)=\left(b_{(i)}\slash n_{(i)}+2\right)\slash \left(b_i\slash n_i+2\right)$ where the subscript $(i)$ denotes \emph{not} $i$.  As in Section~\ref{sec:md}, we now look to maximize and minimize the expression in~\eqref{eq:pgdp}.  As a result of Lemma~\ref{thm:den}, convenient expressions for $C\left(\bx,\bn,\ba,\bb,z_{\cdot}\right)\slash C\left(\by,\bn,\ba,\bb,z_{\cdot}\right)$ do not exist.  Instead, however, we can consider upper and lower bounds on this ratio.
\begin{thm}\label{thm:denrat}
Let $C\left(\by,\bn,\ba,\bb,z_{\cdot}\right)$ be as defined in~\eqref{eq:pg_norm} and let $\bx=\left(x_1,x_2\right)^T$ and $\by=\left(y_1,y_2\right)^T$ denote vectors of non-negative integers of length 2 such that $x_i=y_i-1$ and $x_{i'}=y_{i'}+1$ for $i,i'\in\left\{1,2\right\}$, $i\ne i'$
and $x_1+x_2=y_1+y_2=z_{\cdot}$ and let $\lceil x\rceil^{+}=\max\left(x,0\right)$.  Then when $a_i+y_i < a_{i'}+y_{i'}$,
\begin{align*}
\left\vert\log\frac{C\left(\bx,\bn,\ba,\bb,z_{\cdot}\right)}{C\left(\by,\bn,\ba,\bb,z_{\cdot}\right)}\right\vert \le \left\vert\log\frac{z_{\cdot}\times\lceil 1-r_i\left(\bn,\bb\right)\rceil^{+}+a_{i'} + y_{i'}}{a_i+y_i-1}\right\vert.
\end{align*}
\end{thm}
\noindent
See Appendix~B for a proof for Theorem~\ref{thm:denrat} and an assessment of the bound's accuracy. 

We will now look to maximize the ratio in~\eqref{eq:pgdp} using the result from Theorem~\ref{thm:denrat} assuming (without loss of generality) that $a_1 \le a_2$ and letting $x_1=y_1-1$ and $x_2=y_2+1$.  From Theorem~\ref{thm:denrat}, we have
\begin{align}
\frac{p\left(\bz\given\by,\ba,\bb\right)}{p\left(\bz\given \bx,\ba,\bb\right)} \le&  \frac{z_{\cdot}\times\lceil 1-r_i\left(\bn,\bb\right)\rceil^{+}+a_{2} + y_{2}}{a_1+y_1-1}\times \frac{\left(z_1+a_1+y_1-1\right)}{\left(z_2+a_2+y_2\right)},\label{eq:pgdp1}
\end{align}
which is maximized when $z_1=z_{\cdot}$, $z_2=0$, $y_1=1$, and $y_2=y_{\cdot}-1$, yielding 
\begin{align}
\frac{p\left(\bz\given\by,\ba,\bb\right)}{p\left(\bz\given \bx,\ba,\bb\right)} \le&  \frac{z_{\cdot}\times\lceil 1-r_i\left(\bn,\bb\right)\rceil^{+}+a_{2} + y_{\cdot}-1}{a_2+y_{\cdot}-1}\times \frac{z_{\cdot}+a_1}{a_1}.\label{eq:pgdp2}
\end{align}
Similarly, it can be shown that
\begin{align*}
\frac{p\left(\bz\given\by,\ba,\bb\right)}{p\left(\bz\given \bx,\ba,\bb\right)} \ge&  \frac{a_2+y_{\cdot}-1}{z_{\cdot}\times\lceil 1-r_i\left(\bn,\bb\right)\rceil^{+}+a_2+y_{\cdot}-1}\times \frac{a_1}{z_{\cdot}+a_1},
\end{align*}
and thus if we wish to satisfy $\eps$-differential privacy, we require
\begin{align}
\log \left[ \frac{z_{\cdot}\times\lceil 1-r_1\left(\bn,\bb\right)\rceil^{+}+a_2+y_{\cdot}-1}{a_2+y_{\cdot}-1}\times \frac{z_{\cdot}+a_1}{a_1}\right] = \log \left[\nu_1 \times \frac{z_{\cdot}+a_1}{a_1}\right] \le \eps,\label{eq:pgdp_final}
\end{align}
which implies
\begin{align}
a_i \ge \frac{\nu_i\times z_{\cdot}}{e^{\eps}-\nu_i} = \frac{z_{\cdot}}{e^{\eps}\slash \nu_i -1}\label{eq:pgdp_a}
\end{align}
where $\nu_i$ denotes what amounts to a \emph{penalty} term associated with the additional information gained from using the Poisson-gamma model compared to the multinomial-Dirichlet model.  Note that in practice, $\nu_i \in \left(1,2\right]$ because $z_{\cdot}=y_{\cdot}$ and $a_2\ge 1$.  Furthermore, since we will often consider $a_i$ versus $a_{(i)}$, with $a_i\approx a$ and $a_{(i)}\approx \left(I-1\right) a$ for all $i$, $\nu_i\to 1$ as $a$ increases.  Finally, note that if $b_i\slash n_i=b\slash n$ for some constants $b,n>0$ for all $i$, then the restriction in~\eqref{eq:pgdp_final} is equivalent to the restriction from the multinomial-Dirichlet model in~\eqref{eq:dp_req_md}.  Thus, only modest values of $\eps$ may result in synthetic data that respect the nuances of the true underlying data.

\section{Simulation Study}\label{sec:sim}
To evaluate the performance of the proposed Poisson-gamma synthesizer, we will conduct a simulation study based on the heart disease mortality data described in Section~\ref{sec:data}.  The design of the simulation study and the measures used to assess the performance are described in Section~\ref{sec:sim_design}.  In particular, our focus will be on the extent to which potentially important epidemiologic associations from the true data are retained in the synthetic data.  The results from this comparison are then described in Section~\ref{sec:sim_res}.

\subsection{Underlying data}\label{sec:data}
The dataset used to illustrate the properties of the proposed methodology is comprised of the number of heart disease-related deaths and corresponding population sizes in the counties of the contiguous U.S.\ for those aged 35 and older --- divided into 10-year age groups --- during the year 1980, where deaths due to heart disease are defined as those for which the underlying cause of death was ``diseases of the heart'' according to the 9th revision of the International Classification of Diseases (ICD; ICD--9: 390--398, 402, 404--429).  Because these data are from before the CDC's suppression guidelines \citep{cdc:sharing} went into effect, the public-use data are 
free of suppression and can be obtained via CDC WONDER \citep{data:hd:1980}.  Furthermore, as there were several changes in county definitions during the 1980s, this choice of data from 1980 allows us to use readily-available shapefiles from the Census Bureau for the $I =$ 3,109 counties (or county equivalents) in the contiguous U.S.  Letting $y_{ia}^{\dagger}$ denote the true number of deaths in county $i$ in age group $a$ from a population of size $n_{ia}^{\dagger}$, we then follow the approach of \citet{bym} and \citet{gelfand:mcar} --- i.e., we assume
$y_{ia}^{\dagger} \sim \Pois\left(n_{ia}^{\dagger} \lambda_{ia}\right)$, where $\log \lambda_{ia} \sim \N\left(\beta_{0a} + \phi_{ia},\tau_a^2\right)$,
and $\bphi \sim \MCAR\left(\bG\right)$ --- and consider the posterior medians of the $\lambda_{ia}$ --- denoted $\lambda_{ia}^{\dagger}$ --- as the ``true'' mortality rates for the remainder of our simulation study.

From this point forward, we focus our attention on synthesizing data for the $y_{\cdot 1} = \sum_i y_{i1}^{\dagger} = \text{11,364}$ deaths from those aged 35--44;
additional results are provided in Appendix~C for older age groupings where death counts tend to be higher.  As such, we suppress the age subscript and let $y_{\cdot}$ and $n_{\cdot}$ refer to the number of deaths and total population size for those aged 35--44.

\subsection{Simulation study design and evaluation}\label{sec:sim_design}
To compare and contrast the properties of the differentially private synthesizers described in Section~\ref{sec:methods}, our 
simulation study investigates four scenarios.  In particular, we will 
explore the impact of heterogeneity in the population sizes (yes/no) and heterogeneity in the true underlying event rates (yes/no).  To explore heterogeneity in population sizes, we will compare the results from letting $n_i$ vary according to the age 35--44 population distribution from 1980 --- i.e., $n_i=n_i^{\dagger}$ --- to letting all $n_i = n_{\cdot} \slash I$, the average population size.
To explore heterogeneity in event rates, we will compare results from letting $\lambda_i$ correspond to the posterior medians from the heart disease death data described in Section~\ref{sec:data} --- i.e., $\lambda_i=\lambda_i^{\dagger}$ --- to a scenario in which all $\lambda_i = y_{\cdot}\slash n_{\cdot}$, the mortality rate of the contiguous U.S.
For each scenario, $L=200$ datasets, $\by^{(\ell)}$, will be generated from a Poisson distribution with mean $n_i\lambda_i$ under the constraint that $\sum_i y_{i}^{(\ell)} = y_{\cdot}$.  From each $\by^{(\ell)}$ in each scenario, we will 
use the following synthesis approaches: (a) the multinomial-Dirichlet model, (b) the Poisson-gamma model with smoothing toward the national average, and (c) the Poisson-gamma model with smoothing toward the state averages.  This will be repeated for various levels of $\eps$.

To compare the various approaches, we first recall that in~\eqref{eq:pglik} we assumed $y_i\given \lambda_i \sim \Pois\left(n_i\lambda_i\right)$.  Thus, assuming independence between the $y_i$ (conditional on $\blambda$), the joint distribution of $\by$ conditioned on $\blambda$ and $\sum_i y_i=y_{\cdot}$ is
\begin{align*}
\by\given \blambda,\sum_i y_i=y_{\cdot} \sim \Mult\left(y_{\cdot},\left\{n_i\lambda_i\slash \sum_j n_j \lambda_j\right\}\right).
\end{align*}
That is, $n_i\lambda_i\slash \sum_j n_j \lambda_j$ is the probability of an event occurring in (or in the case of the synthetic data, \emph{being assigned to}) county $i$ under the Poisson-gamma model.  Thus, if we assume $y_{\cdot}\approx \sum_j n_j\lambda_j$, any differences between the synthetic data generated from the Poisson-gamma model and those from the multinomial-Dirichlet model can be attributed to differences between $\lambda_i$ and $\theta_i y_{\cdot}\slash n_i$.  As such, the utility assessment conducted in the simulation study will be done using posterior samples from these parameters rather than the synthetic data themselves. 
In each scenario and for each approach, we will compute the root mean square error (rMSE) --- e.g.,
\begin{align*}
\text{rMSE}\left(\blambda^{(\ell)}\given \blambda\right) = \sqrt{\sum_{i=1}^{I} \left(\lambda_i^{(\ell)}-\lambda_i\right)^2\slash I}
\end{align*}
for $\ell=1,\ldots,L$.  We will multiply the rMSE's by 100,000 (i.e., the typical scale for mortality rates) and present the results with 95\% confidence bands based on the $L$ simulated datasets.  In addition, we will evaluate the urban-rural disparity (where a county is considered ``urban'' if its true 35+ population size is greater than 50,000) and a comparison of the mortality rates in the states of New Mexico and New York.

\subsection{Simulation results}\label{sec:sim_res}
\begin{figure}[t]
    \begin{center}
        \subfigure[Scenario \#1: Same $n_i$; Same $\lambda_i$]{\includegraphics[width=.45\textwidth]{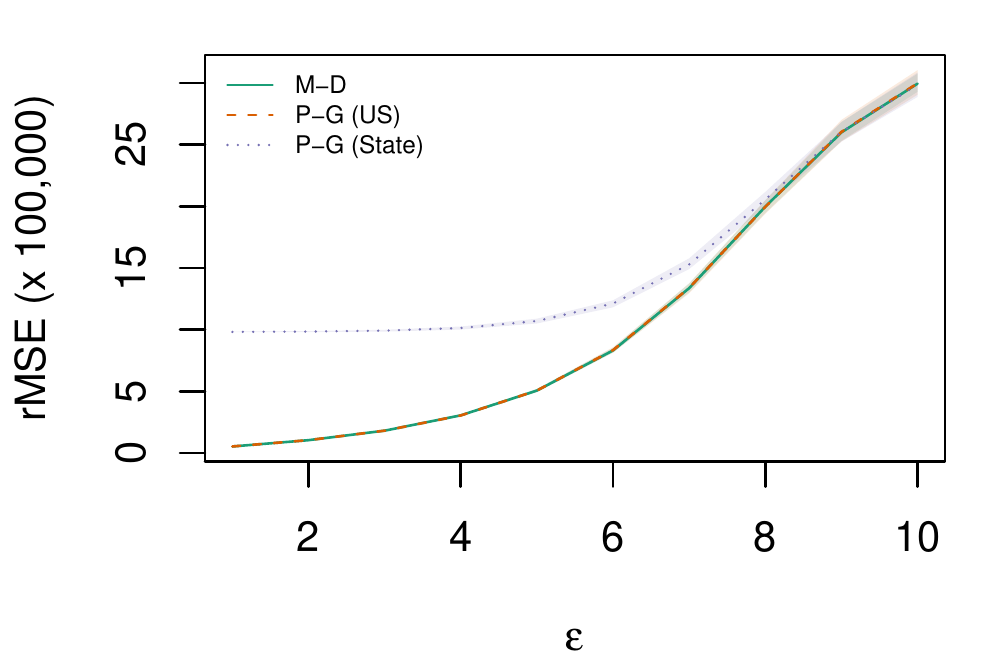}\label{fig:scn1}}
        \subfigure[Scenario \#2: Different $n_i$; Same $\lambda_i$]{\includegraphics[width=.45\textwidth]{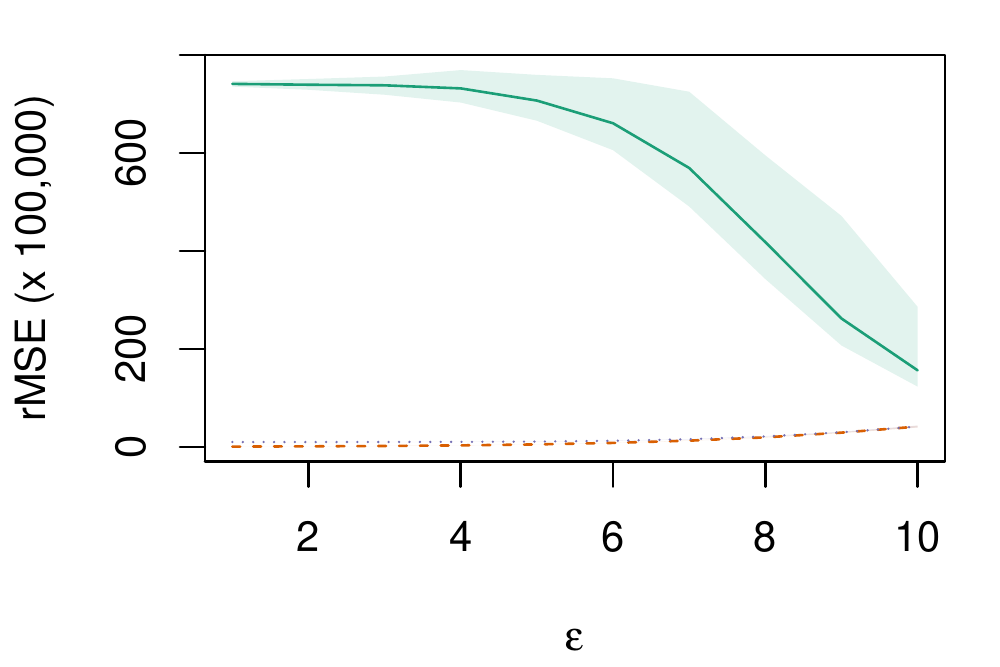}\label{fig:scn2}}\\
        \subfigure[Scenario \#3: Same $n_i$; Different $\lambda_i$]{\includegraphics[width=.45\textwidth]{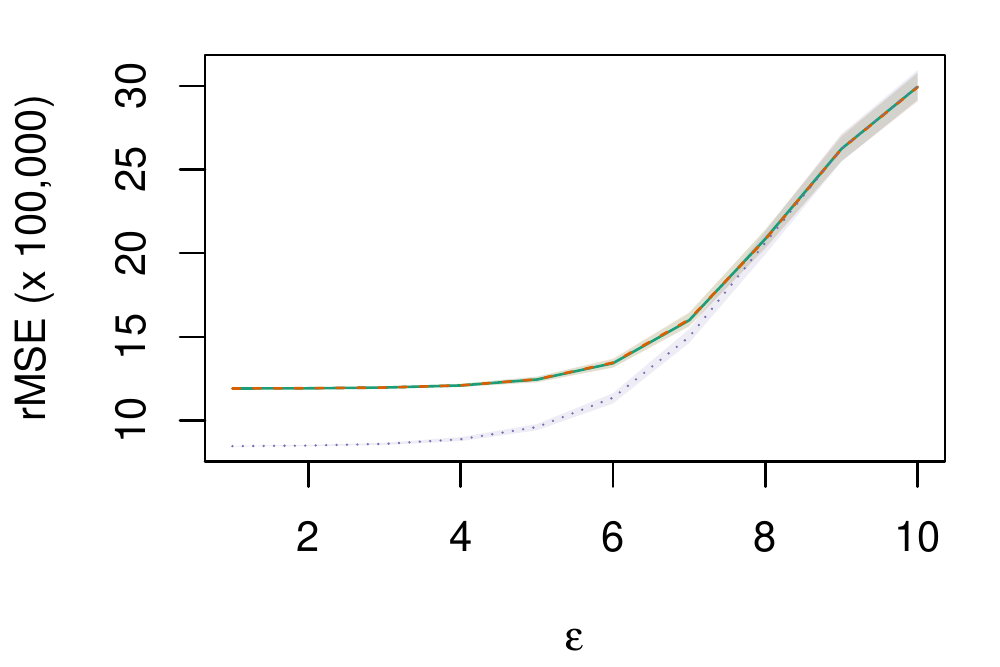}\label{fig:scn3}}
        \subfigure[Scenario \#4: Different $n_i$; Different $\lambda_i$]{\includegraphics[width=.45\textwidth]{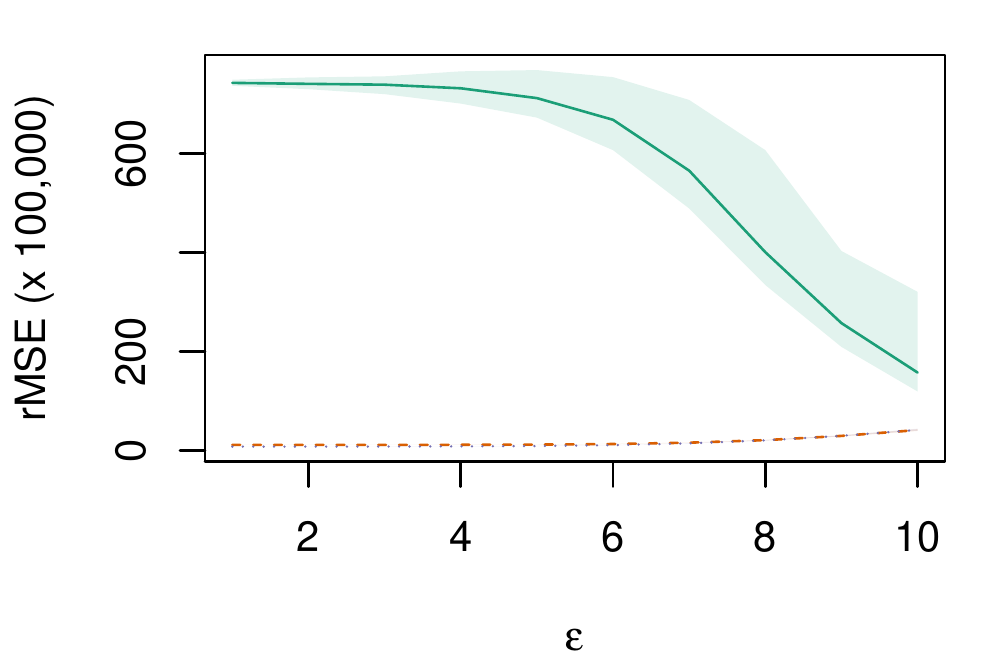}\label{fig:scn4}}
    \end{center}
    \caption{Root mean square error from simulation study.}
    \label{fig:sim}
\end{figure}

Figure~\ref{fig:sim} displays the estimated rMSE's for all four scenarios.  Here, we see that when each county shares the same $n_i$ and $\lambda_i$ (Figure~\ref{fig:scn1}), the multinomial-Dirichlet model and the Poisson-gamma model that smooths toward the national average yield equivalent results and both slightly outperform the Poisson-gamma approach that smooths toward the state-specific averages as $\eps$ decreases.  Similarly, when the population sizes are the same but the $\lambda_i$'s are allowed to vary (Figure~\ref{fig:scn3}), the edge goes to the Poisson-gamma approach that smooths toward the state-specific means by a small margin.  When $n_i$ is allowed to vary, however, the rMSE's from both of the Poisson-gamma models dominate those from the multinomial-Dirichlet model, even for large $\epsilon$.

In retrospect, it is clear why this occurs.  When all counties share the same population size, the multinomial-Dirichlet model with $\alpha_i=\alpha$ is mathematically \emph{equivalent} to the Poisson-gamma model with $\lambda_i=y_{\cdot}\slash n_{\cdot}$ and $a_i=a$.  To see why the multinomial-Dirichlet model performs poorly compared to the Poisson-gamma models when the population sizes vary, we look to the expected values for the synthetic data.  Under the multinomial-Dirichlet approach from Section~\ref{sec:md},
\begin{align}
E\left[z_i\given\by,\balpha\right] &= E\left(E\left[z_i\given \btheta\right]\given \by,\balpha\right)= E\left[\theta_i z_{\cdot}\given \by,\balpha\right] = \frac{y_i+\alpha_i}{\sum_j y_j+\alpha_j} z_{\cdot} \to \frac{\alpha_i}{\sum_j \alpha_j} z_{\cdot} = \frac{z_{\cdot}}{I}; \label{eq:mdmean}
\end{align}
i.e., as $\alpha_i\to \infty$, the events will become uniformly distributed among the $I$ counties.  In contrast, if $z_{\cdot}=y_{\cdot}$ and $\sum_j n_j\lambda_j \approx y_{\cdot}$, the expected value for $z_i$ under the Poisson-gamma model from Section~\ref{sec:pg} yields
\begin{align}
E\left[z_i\given \by,\ba,\bb\right] &= E\left(E\left[z_i\given \blambda\right]\given \by,\ba,\bb\right) = E\left[\frac{n_i\lambda_i}{\sum_j n_j\lambda_j}z_{\cdot}\given \by,\ba,\bb\right]\approx \frac{y_i+a_i}{n_i+b_i} n_i\to n_i\lambda_{0i}; \label{eq:pgmean}
\end{align}
i.e., as $a_i,b_i\to \infty$, the events will be distributed in a manner which reflects the population sizes and prior event rates, $a_i\slash b_i=\lambda_{0i}$, of the counties. That is, when there is heterogeneity in the population sizes, we should expect the Poisson-gamma model to produce synthetic data with greater utility than the multinomial-Dirichlet model with an equivalent risk of disclosure.  Furthermore, when we allow the model to use existing prior information regarding heterogeneity in the event rates, additional gains in utility should be expected.

\begin{figure}[t]
    \begin{center}
        \subfigure[Urban versus Rural]{\includegraphics[width=.45\textwidth]{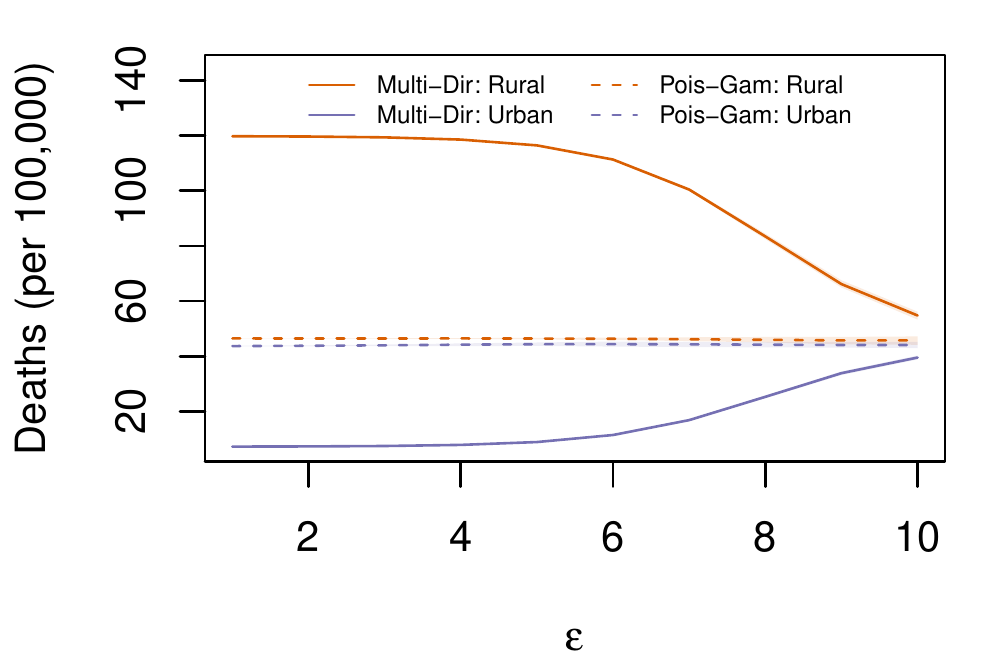}\label{fig:urbanrural}}
        \subfigure[New Mexico versus New York]{\includegraphics[width=.45\textwidth]{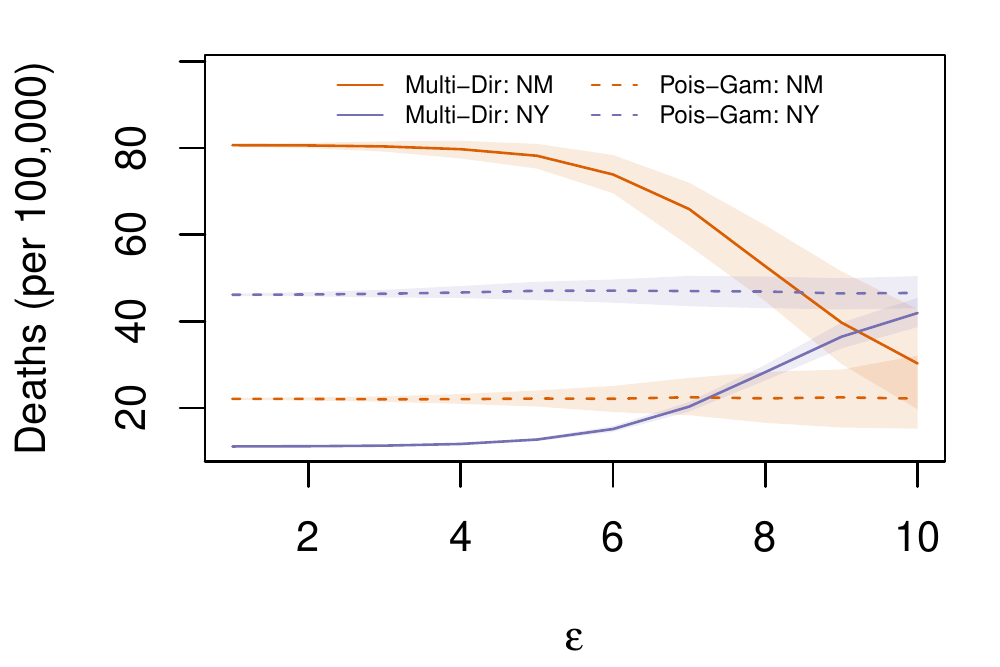}\label{fig:nmny}}
    \end{center}
    \caption{Illustration of properties of the utility of the synthetic data.  Synthetic data generated from the Poisson-gamma model are smoothed toward the state-specific averages.}
    \label{fig:illustration}
\end{figure}

To put these results in context, we consider the extent to which the urban-rural disparity and comparison of rates in New Mexico and New York can be distorted by the data synthesis process under Scenario~4.
In Figure~\ref{fig:urbanrural}, we compare the heart disease mortality rate of urban counties to that from rural counties.  Because there are far more rural counties than urban counties, the multinomial-Dirichlet model allocates a disproportionate number of deaths to rural counties, thus dramatically inflating their rates.  In contrast, the Poisson-gamma model with smoothing toward state-specific averages produces rate estimates for both urban and rural counties that are on par with the truth.
Similarly, Figure~\ref{fig:nmny} displays a comparison of estimated rates for New York versus New Mexico.  Because New York has 14 times the adult population of New Mexico and has only 2 times as many counties, the multinomial-Dirichlet model produces rate estimates for New Mexico that are approximately $14/2=7$ times greater than those in New York.  This is again in contrast to the Poisson-gamma model with smoothing toward the state-specific averages, which (by design) produces accurate estimates for all $\eps$.
Thus, while synthetic data produced by the Poisson-gamma model may yield more conservative inference (particularly when smoothing toward the national average), synthetic data from the multinomial-Dirichlet model may yield estimates that exhibit both Type-M (magnitude) and Type-S (sign) errors \citep{gelman:ms}.

\section{Discussion}\label{sec:disc}
In this paper, we have generalized the approach of \citet{onthemap} for generating differentially private synthetic data from the multinomial-Dirichlet setting to a more flexible Poisson-gamma setting.  In addition to decomposing a collection of abstract probability parameters into a function of interpretable offset (e.g., population size) and rate parameters, the Poisson-gamma setting also grants data stewards more control over the utility of the synthetic data. 
As we have demonstrated via simulation and proved mathematically, the Poisson-gamma approach can be equivalent to the multinomial-Dirichlet approach in the most simple of settings (equal population sizes and prior beliefs) while its added flexibility can yield far greater utility in more realistic settings. 

One note that we have glossed over to this point is the notion that this approach assumes that (and conditions on) certain pieces of information are safe to be disclosed.  In particular, we have assumed in our simulation study that the total number of deaths due to a certain cause of death in a particular age group is known (i.e., $y_{\cdot}$).  This implies that we are comfortable with an intruder knowing that a given individual died of this cause of death at a certain age but not which \emph{county} the individual lived in.  In other settings, however, an agency may assume that demographic attributes such as an individual's age, race, sex, and county are known, but that their \emph{cause of death} is unknown.  In this scenario, the presentation described here could simply be reframed to synthesize the number of individuals assigned to each ICD-10 code within these demographic strata -- because the total number of deaths occurring in each strata would likely be small, differential privacy could be satisfied with relatively noninformative priors even for small $\epsilon$.  This provides yet another lever for agencies to manipulate to improve the utility of the data they release.

Another avenue for improving the utility of the synthetic data produced by the proposed Poisson-gamma model would be to consider the $\left(\eps,\delta\right)$-probabilistic differential privacy framework proposed by \citet{onthemap}.  As described in Section~\ref{sec:md}, satisfying $\eps$-differential privacy requires bounding the maximal risk, which occurs when $y_i=1$ in the true data and all $z_{\cdot}$ events are assigned to the $i$th cell in the synthetic data.  Because the probability of sampling such an extreme synthetic dataset is very small, \citet{onthemap} proposed an algorithm for defining $\balpha$ such that the probability of sampling a synthetic dataset, $\bz$, that violates~\eqref{eq:dp_dfn} was less than a user-defined $\delta>0$.
While it is possible to derive a similar framework for the Poisson-gamma model proposed here, we believe an approach that \emph{truncates} the range of each $z_i$ based on its \emph{prior predictive distribution}, $p\left(z_i\given a_i,b_i\right)$, could be more intuitive to implement 
and result in a true differentially private mechanism (i.e., $\delta=0$). 



Finally, we would be remiss to not discuss the relationship between the proposed Poisson-gamma model and more conventional approaches for satisfying differential privacy --- i.e., those that sanitize the truth by adding differentially private noise.  In particular, one might suspect that such approaches would struggle when there is substantial heterogeneity in the counts
because the impact of adding noise to $y_i=0$ events from a small $n_i$ 
is much different than adding the same amount of noise to a large $y_{i'}$ out of a similarly large $n_{i'}$ 
--- i.e., the noise added is not itself a function of the number of events or the population size.  One compromise might be the model-based differentially private synthesis (MODIPS) approach of \citet{liu:modips} in which posterior distributions are sanitized by adding differentially private noise to the sufficient statistics.  In particular, the approach of \citet{liu:modips} could easily be extended to the Poisson-gamma setting in which the posterior distribution in~\eqref{eq:lambdapost} would be replaced by
\begin{align}
\lambda_i^*\given y_i \sim \Gam\left(a_i+f\left(y_i\right),b_i+n_i\right),\label{eq:modipspost}
\end{align}
where $f\left(y_i\right)= y_i+e_i$ and $e_i$ is differentially private noise.  While the $a_i+e_i$ in~\eqref{eq:modipspost} will surely be less than the $a_i$ required to satisfy $\eps$-differential privacy in~\eqref{eq:pgdp_a} --- suggesting the potential for improved utility --- the drawback of~\eqref{eq:modipspost} is that $E\left[\lambda_i^*\given y_i\right] = \left(a_i+y_i+e_i\right)\slash \left(b_i+n_i\right)$ is not a weighted average the crude rate from the data, $y_i\slash n_i$, and the prior expected value, $a_i\slash b_i$. As a result, the benefit of using less informative priors could potentially be negated by smoothing $\lambda_i^*$ toward 
$\left(a_i+e_i\right)\slash b_i$.
Alternatively, we could
sanitize the \emph{gamma prior} in~\eqref{eq:pglik} itself.  For instance,
instead of smoothing the $\lambda_i$ toward their state-specific averages --- $\lambda_{0;s_i} = \sum_{j\in {\cal S}_{s_i}} y_j \slash \sum_{j\in {\cal S}_{s_i}} n_j$, where ${\cal S}_s$ denotes the set of counties belonging to state $s$ --- we could let $b_i^* = a_i\slash \lambda_{0;s_i}^*$ and $\lambda_i \sim \Gam\left(a_i,b_i^*\right)$ where
\begin{align}
\lambda_{0;s_i}^* = \frac{\sum_{j\in {\cal S}_{s_i}} y_j + e_{s_i}}{\sum_{j\in {\cal S}_{s_i}} n_j}\label{eq:alt}
\end{align}
and $e_{s_i}$ is differentially private noise.
A key benefit of this approach would be that the noise in~\eqref{eq:alt} would be added to larger, aggregate counts, thereby resulting in less degradation to the utility of the model.  Nevertheless, the focus of this paper is not to argue which approach is optimal (a status that is likely application specific),
but rather to demonstrate the conditions in which the Poisson-gamma model satisfies differential privacy.

Issues with the utility of differentially private synthetic data are not new.  In particular, \citet{charest:2010} highlighted the inherent bias in~\eqref{eq:mdmean} for the multinomial-Dirichlet model due to the tendency to let $\alpha_i =\alpha$ for all $i$.  That is, unlike in most Bayesian statistical approaches for generating synthetic data, prior distributions in differentially private synthesizers tend to be designed \emph{solely} to satisfy differential privacy for a certain $\eps$ rather than represent one's prior beliefs or best capture the data's complex dependence structures.  To overcome this bias, \citet{charest:2010} proposed what essentially amounts to a measurement error model 
in which the synthetic data are treated as a noisy version of the truth and the end-user attempts to estimate the truth from the synthetic data --- i.e., make inference on $p\left(\by\given\bz,\balpha\right)$.
The drawback of this approach, however, is that it assumes (a) that end-users are aware of this bias, (b) that end-users are savvy enough to do such preprocessing of the public-use data, and (c) that agencies would disclose the details of their data synthesizers --- including the level of $\eps$ used --- to accurately recover or approximate the true data.

While such techniques can be effective for overcoming the bias induced by sanitizing data for public-use, the proposed work is intended as a step toward differentially private synthetic data that require no preprocessing on the part of would-be data users. 
Specifically, the Poisson-gamma model has parameters to control the informativeness of the model, $\ba$, and parameters that dictate what the model smooths estimates toward, $\bb$.  As illustrated here, this framework allowed us to account for heterogeneity in both population sizes and prior event rates to yield synthetic data with substantially improved utility.  In our future work, we aim to develop further extensions of formal privacy guarantees to nonconjugate models, thereby permitting the creation and dissemination of differentially private synthetic data that benefit from more conventional spatial and spatiotemporal model structures like those used by \citet{quick:synthetic}.  In the meantime, we believe that utility can be improved via truncating synthetic counts to reasonable ranges and stratification; e.g., synthesizing counts
based on demographics such as age group, race/ethnicity, and sex and by geographic regions like the Census Regions and Divisions.

\bibliographystyle{jasa}
\bibliography{cdc_ref,cdc_epi,reports,disclosure,wonder}

\end{document}